# Inverse-designed nanophotonic neural network accelerators for ultra-compact optical computing

Joel Sved[1,2], Shijie Song[1,2], Liwei Li[1,2], George Li[1,2], Debin Meng[1,2] & Xiaoke Yi[1,2]*

**Author affiliations.** [1]School of Electrical and Computer Engineering, the University of Sydney, NSW 2006, Australia.
[2]Sydney Nano Institute, The University of Sydney, NSW 2006, Australia.
*E-mail: xiaoke.yi@sydney.edu.au

## Abstract

Inverse-designed nanophotonic devices offer promising solutions for analog optical computation. High-density photonic integration is critical for scaling such architectures toward more complex computational tasks and large-scale applications. Here, we present an inverse-designed photonic neural network (PNN) accelerator on a high-index contrast material platform, enabling ultra-compact and energy-efficient optical computing. Our approach introduces a wave-based inverse-design method based on three dimensional finite-difference time-domain (3D-FDTD) simulations, exploiting the linearity of Maxwell's equations to reconstruct arbitrary spatial fields through optical coherence. By decoupling the forward-pass process into linearly separable simulations, our approach is highly amenable to computational parallelism, making it particularly well suited for acceleration using graphics processing units (GPUs) and other parallel computing platforms, thereby enhancing scalability across large problem domains. We fabricate and experimentally validate two inverse-designed PNN accelerators on the silicon-on-insulator platform, achieving on-chip MNIST and MedNIST classification accuracies of 89% and 90% respectively, within ultra-compact footprints of just $20 \times 20$ μm² and $30 \times 20$ μm². Our results establish a scalable and energy-efficient platform for analog photonic computing, effectively bridging inverse nanophotonic design with high-performance optical information processing.

## Introduction

Machine learning has transcended conventional signal analysis techniques in numerous research fields ranging from computer vision to natural language processing[1,2]. Central to this progress is the ability of neural networks to learn complex data representations, enabling robust and adaptive inference. However, as neural models continue to grow in complexity, conventional electronic hardware faces increasing challenges in meeting the demands of computing speed and energy consumption[3-6]. In response, photonic neural networks (PNNs) have emerged as a promising platform for analog neural computation, offering key advantages such as ultrafast processing and low-power consumption[4-9].

To unlock the full potential of PNNs, it is essential to achieve high-density integration capable of supporting complex models and large datasets on-chip[3,6,10]. Traditional photonic design approaches, while effective for standard components, often fall short when optimizing for compactness, multifunctionality, and performance simultaneously. In this context, topology optimization based inverse-design methods have emerged as a powerful tool[11-14]. By computationally exploring a vast design space unconstrained by human intuition, it enables the discovery of non-intuitive geometries that maximize light-matter interaction within a minimal footprint[15]. Unlike diffractive models[4-6], which rely on sequential, layer-by-layer architectures mimicking feedforward deep neural networks (DNNs), topology optimization allows arbitrary control over scattering, interference, and phase evolution throughout the device's entire volume, enabling more compact and functionally dense implementations[16-18].

However, the practical deployment of topology-optimized PNNs to accommodate increased dataset sizes remains challenging due to the absence of a scalable, physics-accurate training framework. Moreover, increasing the refractive index contrast enhances signal expressivity[19], by strengthening confinement, supporting internal resonances[20], and enabling stronger interference and engineered scattering effects[21] - all of which are critical for high-fidelity optical computation. Consequently, three dimensional solvers such as three dimensional finite-difference time-domain (3D-FDTD) are required, since approximation methods such as the effective index method often fail to accurately capture the resulting out-of-plane scattering dynamics[21-23]. Moreover, as neural architectures scale to support increasingly complex tasks and larger datasets, computational efficiency and scalability become essential. Thus, achieving ultra-compact PNNs suitable for dense on-chip integration necessitates a design approach that captures the full-wave physics of high-index-contrast nanophotonic structures, while remaining computationally scalable and fabrication feasible.

Here, we demonstrate ultra-compact, inverse-designed PNN accelerators fabricated on a high-index-contrast silicon-on-insulator (SOI) platform operating at a single wavelength. To enhance the inverse-design efficiency, we develop a wave-based inverse-design approach grounded in 3D-FDTD simulations, which exploits the linearity of Maxwell's equations to reconstruct arbitrary spatial fields through optical coherence. By decoupling the forward-pass process into linearly independent simulations, our approach enables efficient parallelization and scales naturally across high-throughput computing hardware, such as graphics processing units (GPUs). The fabricated devices, with footprints as small as $20 \times 20$ µm² and $30 \times 20$ µm², perform on-chip image classification, demonstrating their efficacy on benchmarking datasets (MNIST[24] and MedNIST[25-27]), and achieving experimental accuracies of 89% and 90% respectively. Our inverse-designed PNN demonstrates the practical feasibility of ultra-compact nanophotonic devices for accelerating on-chip data processing. Owing to the method's scalability and inherent compatibility with parallel computing architectures, this work is extensible to higher-dimensional processing tasks, offering a scalable pathway to efficiently handle increasingly demanding computational workloads.

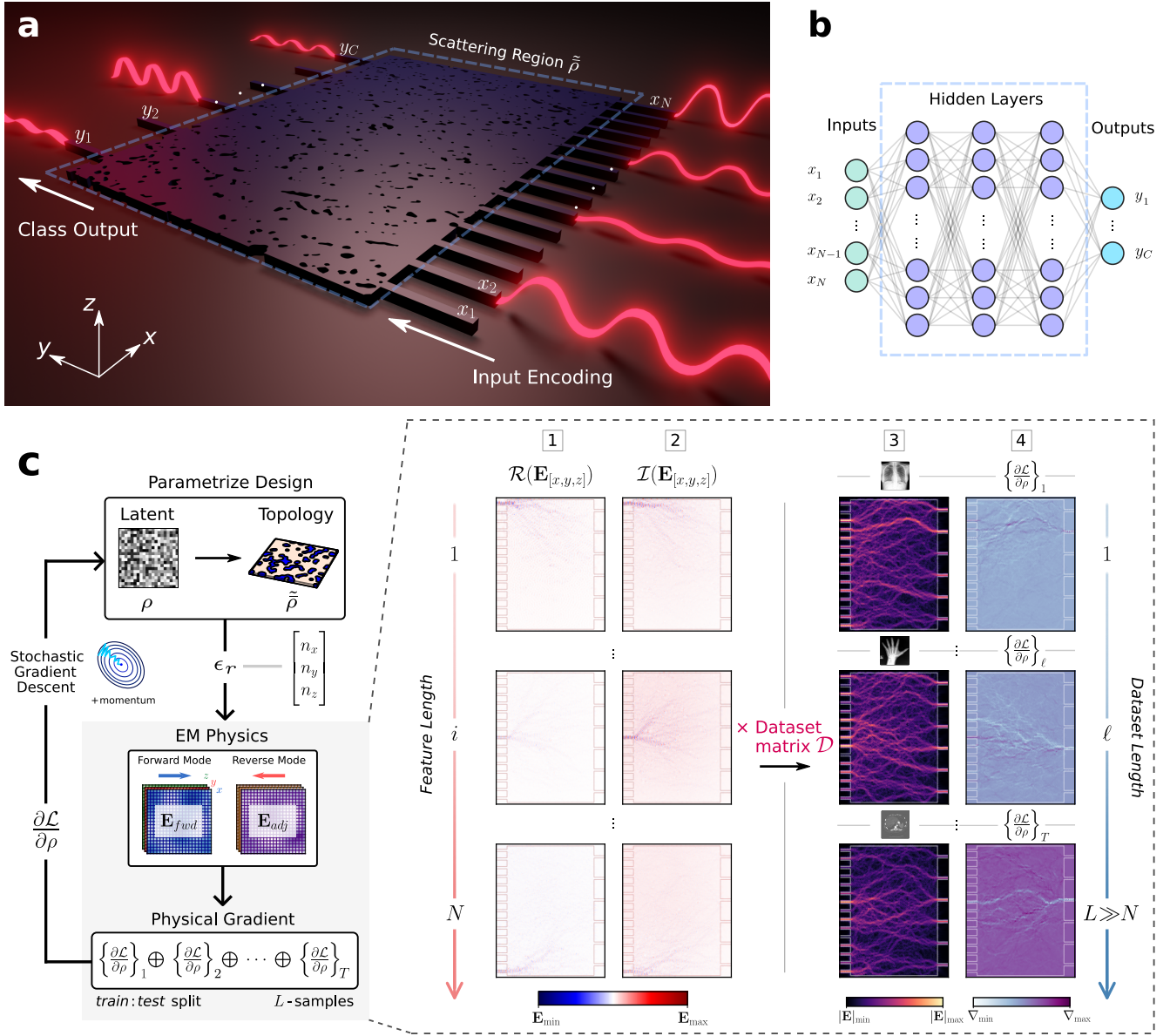

**Fig. 1 Inverse-designed nanophotonic neural network accelerator. a** Schematic diagram of proposed on-chip PNN accelerator with $N$ input feature encoding $x_1, x_2, \ldots x_N$, $C$ class output distribution $y_1, y_2, \ldots y_C$ and scattering region $\tilde{\rho}$. **b** Corresponding schematic of an analogous digital multi-layer perceptron network. **c** Flow chart detailing the inverse-design and training procedure for on-chip PNN. A latent design feature space $\rho$ is parametrized into a real permittivity matrix $\epsilon_r$ of dimension $[n_x, n_y, n_z]$, representing spatial discretization in the $x$, $y$ and $z$ directions. Subsequently, complex **E**-fields are generated, and a physics-based gradient is computed via AVM for backpropagation. This gradient is derived from the $L$-forward fields ($\mathbf{E}_{fwd}$) and $T$-reverse mode fields ($\mathbf{E}_{adj}$). Inset summaries the inverse-design forward-pass and backpropagation procedures. Columns 1-2 illustrate the real $\mathcal{R}(\cdot)$ and imaginary $\mathcal{I}(\cdot)$ components of the $N$-coherent forward mode fields $\mathbf{E}_{[x,y,z]}$ ($x$, $y$, $z$ superimposed) at indexes $1, \ldots, i, \ldots, N$ with values bounded between $[\mathbf{E}_{min}, \mathbf{E}_{max}]$. Column 3 shows the field magnitude $|\mathbf{E}|$ for exemplary training samples in classes 'CXR', 'Hand' and 'AbdomenCT' at indexes $1, \ldots, \ell, \ldots, T$ from matrix multiplication with dataset feature matrix $\mathcal{D}$. Importantly, this operation reconstructs a full set of $L$ simulated fields from only a limited $N$ number of forward simulations. Column 4 captures the corresponding $T$-reverse mode induced gradient $\nabla = \partial \mathcal{L}/\partial \rho$ spatial fields which are concatenated ($\oplus$) and used to optimize the parameter space with respect to the cross-entropy loss function $\mathcal{L}$ with values bounded between $[\nabla_{min}, \nabla_{max}]$.

**Results**

**PNN accelerator configuration and inverse-design.** Figure 1a presents a schematic overview of the on-chip PNN accelerator, wherein compressed and flattened input features from the dataset are encoded onto coherent optical amplitudes at a single wavelength. Within the topology-optimized scattering region, the encoded optical fields undergo complex interference and scattering interactions throughout the device volume. This design enables non-uniform and spatially adaptive transport of optical energy, efficiently routing optical power from the $N$ input encoding ports toward the $C$ class output ports. The PNN outputs represent the probability distribution across classification categories, analogous to the output layer of a conventional multi-layer perceptron network (Fig. 1b).

To optimize the device material index distribution across the scattering region, we directly perform the optimization within the binarized regime, enforcing fabrication constraints at every design epoch (see **Methods**). This approach removes the necessity for hyperparameter scheduling and post-processing discretization[28]. As illustrated in Fig. 1c, the latent design feature space, which varies continuously between 0 and 1, undergoes low-pass filtering to smooth parameter spatial distributions and is subsequently binarized[29] into fabrication-compatible materials: silicon (Si) and silicon dioxide ($SiO_2$) as representative examples. We apply morphological operations to remove small features[30], and B-spline contour approximations[31–33] to further refine and shape the final geometry under minimal feature size and radius of curvature constraints, ensuring manufacturability[34].

To accurately capture the performance of our design applied to multi-class image classification, it is suitable to minimize the cross-entropy (CE)[1] loss between the one-hot encoded ground truth labels and the output optical power vector. This objective function indicates degraded performance when the optical power at the target output port is low. The corresponding output probability distribution across classes is computed by taking the surface integral of the optical power density at each port normalized by their total sum[17]. As a result of this normalization, the CE loss function implicitly penalizes higher distributed power at non-target ports, which is important for limiting channel crosstalk. The gradient of the loss function with respect to the full design space is computed using the adjoint variable method (AVM)[35,36] and is proportional to the overlap of the forward mode fields and the adjoint fields. For each classification category, an adjoint source is constructed from the one-hot encoded target vector, which is used to derive the reverse-mode field ($\mathbf{E}_{adj}$).

To simulate the forward-pass optical fields, we exploit the linearity of Maxwell's equations to significantly reduce computational complexity. Rather than individually simulating each input sample across a dataset of size $L$ using 3D-FDTD, we only execute a limited number ($N$, the number of input encoding ports) of forward-mode 3D simulations, as illustrated in columns 1 and 2 of the inset in Fig. 1c. Leveraging the coherent nature of optical fields[37], one can derive sample-wise fields ($\mathbf{E}_{fwd}^{\ell}$) as a linear combination of precomputed mode source fields $\mathbf{E} = [\mathbf{E}_1, \mathbf{E}_2, ..., \mathbf{E}_N]$, such that $\mathbf{E}_{fwd}^{\ell} = \sum_{i=1}^{N} a_i^{\ell} \mathbf{E}_i$, where $a_i^{\ell}$ are the coefficients mapped to dataset features at sample $\ell$ and $\mathbf{E}_i$ is the full 3D field from the $i$-th mode source. More generally, the entire dataset of reconstructed optical fields can be compactly expressed as $\mathbf{E}_{fwd} = \mathcal{D}^T \mathbf{E}$, where $\mathcal{D} \in \mathcal{R}^{N \times L}$ is the dataset feature matrix encoding all input samples, and $\mathbf{E}_{fwd}$ is the tensor containing the corresponding forward fields. The information is reshaped into a complex tensor with dimensions $[L, n_x, n_y, n_z]$, where $n_x$, $n_y$ and $n_z$ denote the matrix dimensions in the $x$, $y$ and $z$-axes respectively. From this tensor, a subset $T \subset L$ is selected as training fields, which directly inform the iterative updates during inverse-design optimization. Representative training fields corresponding to sample images from the MedNIST dataset are shown in Fig. 1c, inset column 3.

The exact gradient update, as illustrated in Fig. 1c inset, column 4 is obtained by solving the AVM equation across the training samples. This gradient acquisition differs from straight-through estimators which approximate the physical gradient through binary activation functions using an identity operator - a process which is a commonly used in training quantized DNNs[38]. Using stochastic gradient descent with momentum (Fig. 1c), we iteratively update the latent parameter distribution to minimize the CE loss, thus progressively increasing the classification accuracy during optimization (see **Methods**).

The total number of required 3D-FDTD simulations per epoch is only $N + C$, composing $N$ simulations (forward mode) to precompute input-related fields, plus $C$ simulations (reverse mode) to compute gradients for each class category. This approach substantially lowers computational costs compared to directly simulating the entire dataset, especially beneficial when dealing with large datasets, where the total dataset size $L$ is significantly greater than $N + C$[2,39]. Furthermore, by virtue of the linear separability of the forward mode spatial fields, this inverse-design approach is highly amenable to computational parallelism, making it particularly well suited for acceleration on GPU and other parallel computing architectures, thereby enhancing scalability across large problem domains.

**On-chip image classification tasks.** To demonstrate the scalability of our inverse-design scheme applied to image classification, we perform numerical experiments using two benchmark datasets with progressively increased complexity: the MNIST dataset, containing 70,000 grayscale digit images[24] with 10 input features ($N = 10$), 10 classes ($C = 10$) and a design region footprint $20 \times 20$ μm$^2$, and the MedNIST containing 58,954 medical images[25-27] with 15 features ($N = 15$) across six classes ($C = 6$) and a design region footprint of $30 \times 20$ μm$^2$. For both datasets, the input data features are spatially encoded onto the amplitude of light at a wavelength of 1550 nm. The MNIST dataset is partitioned into 50,000 training images and 20,000 test images, while the MedNIST dataset includes 48,000 training images and 10,954 test images. Given these datasets are in grayscale, it is only necessary to restrict the problem domain to the time-harmonic regime in numerical simulations[15].

The MNIST classification results are shown in Fig 2. Figure 2a illustrates the normalized power distribution across all output ports and the corresponding electric field (**E**−field) propagation at epochs 0, 60, and 120 for a randomly selected example from Class '5'.

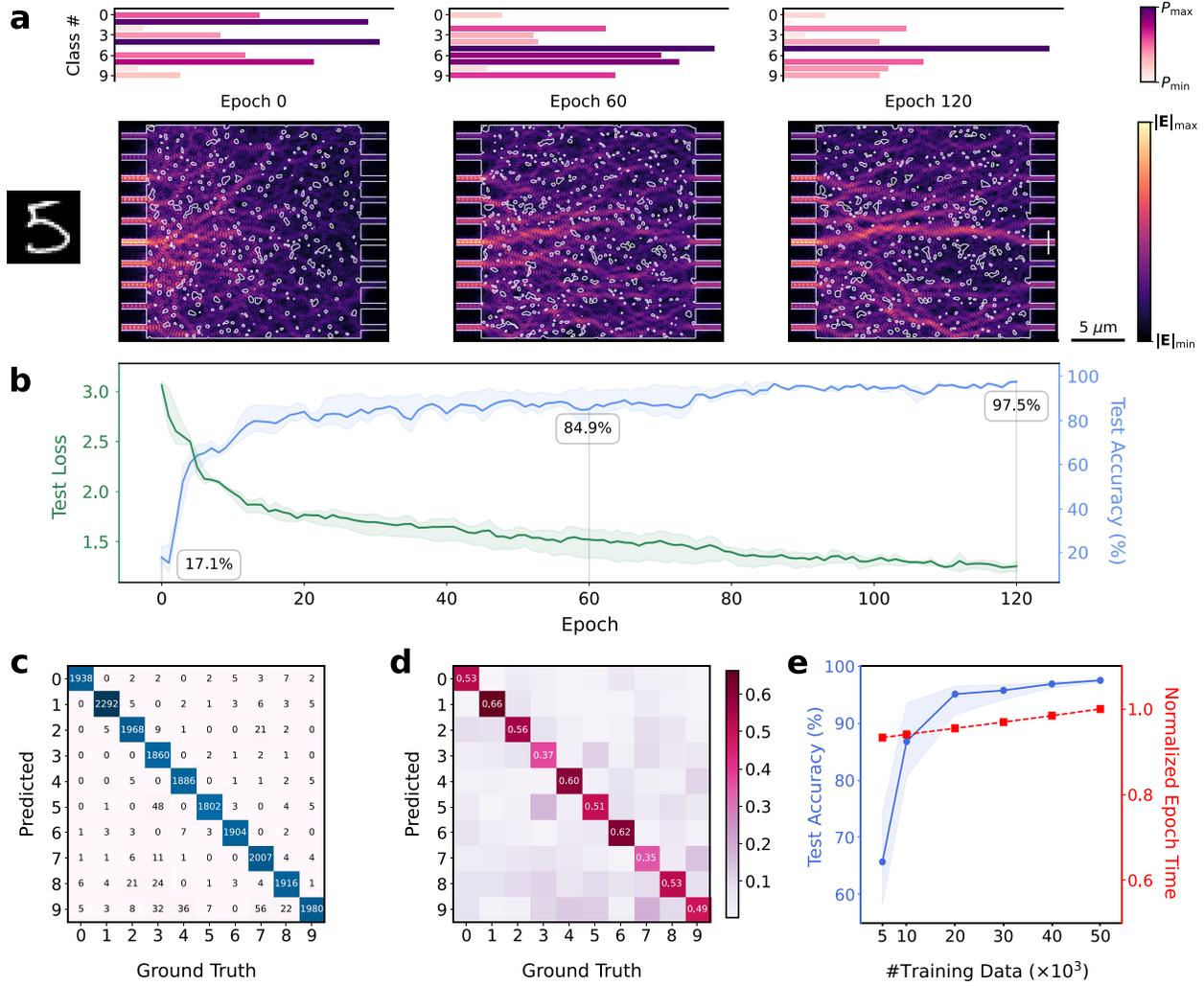

**Fig. 2 MNIST 10-class classification results**. **a** Normalized optical power distribution $P$ with respect to a maximum value $P_{max}$ and minimum value $P_{min}$ across output ports at epochs 0, 60 and 120, shown alongside the corresponding **E**-field magnitude maps at center-line, with maximum and minimum values denoted as $|\mathbf{E}|_{min}$ and $|\mathbf{E}|_{max}$ respectively for a randomly selected test sample from class '5'. Discrete index contours highlight the evolving hot-spot concentration at the correct output port. **b** Test loss and accuracy curves as a function of epochs with range overlayed across three random data permutations. Green: Test CE loss. Blue: Test classification accuracy. **c** Confusion matrix for 10-class image classification. **d** Energy density matrix at final epoch. **e** Blue: Mean test accuracy as a function of training data, with range overlaid across three random data permutations. Red: Max-normalized time taken for single training epoch as a function of training data.

Initially, the randomized high-contrast material topology does not yield meaningful classification outcomes. However, as topology optimization progresses, optical energy becomes increasingly concentrated at the correct output port corresponding to Class '5', while power at neighbouring ports are effectively suppressed. Figure 2b depicts test loss and accuracy as a function of training epochs, with range intervals capturing variability over three separate training sequences where the datasets are randomly permutated. The final confusion matrix (Fig. 2c) demonstrates a high overall classification accuracy of 97.8%. In addition, the energy density matrix (Fig. 2d), as a measure of the output power distribution averaged across the test set, confirms effective energy localization, showing the optical energy is well concentrated at the correct output ports corresponding to each class, thereby indicating minimal cross-class optical crosstalk. Moreover, Fig. 2e illustrates the classification accuracy with respect to training data size, reinforcing the need for high volumes of training data to perform accurate signal classification. Notably, despite the increase in training data volume, our method maintains a constant number ($N + C = 20$) of required 3D-FDTD simulations per epoch in the inverse-design. As a result, the increase in epoch runtime is just 6.7% when using the training full dataset compared to only 10% of it.

Next, we perform classification on the MedNIST dataset to evaluate the scalability of the proposed method to a higher-dimensional dataset with a larger volume of data. Figure 3a shows the inference fields for a single randomly selected sample from each of the six MedNIST classes, along with their corresponding normalized power distributions, demonstrating correct signal localization at the respective class output port. As shown in Fig. 3b, the optimizer reached an equilibrium after 150 epochs, yielding a peak numerical accuracy of 99.1%. The optimized material topology at the final training epoch is shown in the 3D-rendered schematic (Fig. 3b Inset), highlighting the 220 nm fully-etched Si waveguide designed to enable efficient wave propagation through its high refractive-index contrast. The confusion matrix (Fig. 3c) quantitatively confirms this high classification accuracy, reaching 100% accuracies for classes 'ChestCT', 'AbdomenCT', 'HeadCT' and 'BreastMRI'. Furthermore, the energy density matrix (Fig. 3d) demonstrates effective energy localization, with approximately half of the total output power concentrated at the correct output port for five classes. Although slight coupling to neighboring ports is observed, the device maintains high performance even within its ultra-compact footprint. Figure 3e further validates performance of our 3D based optimization approach when volume training data increases. The observed classification accuracy improves consistently with increasing dataset size, while the computational overhead remains minimal due to the fixed number of FDTD simulations per epoch.

The results underscore the scalable nature of our inverse-designed PNN in processing high-volume complex data.

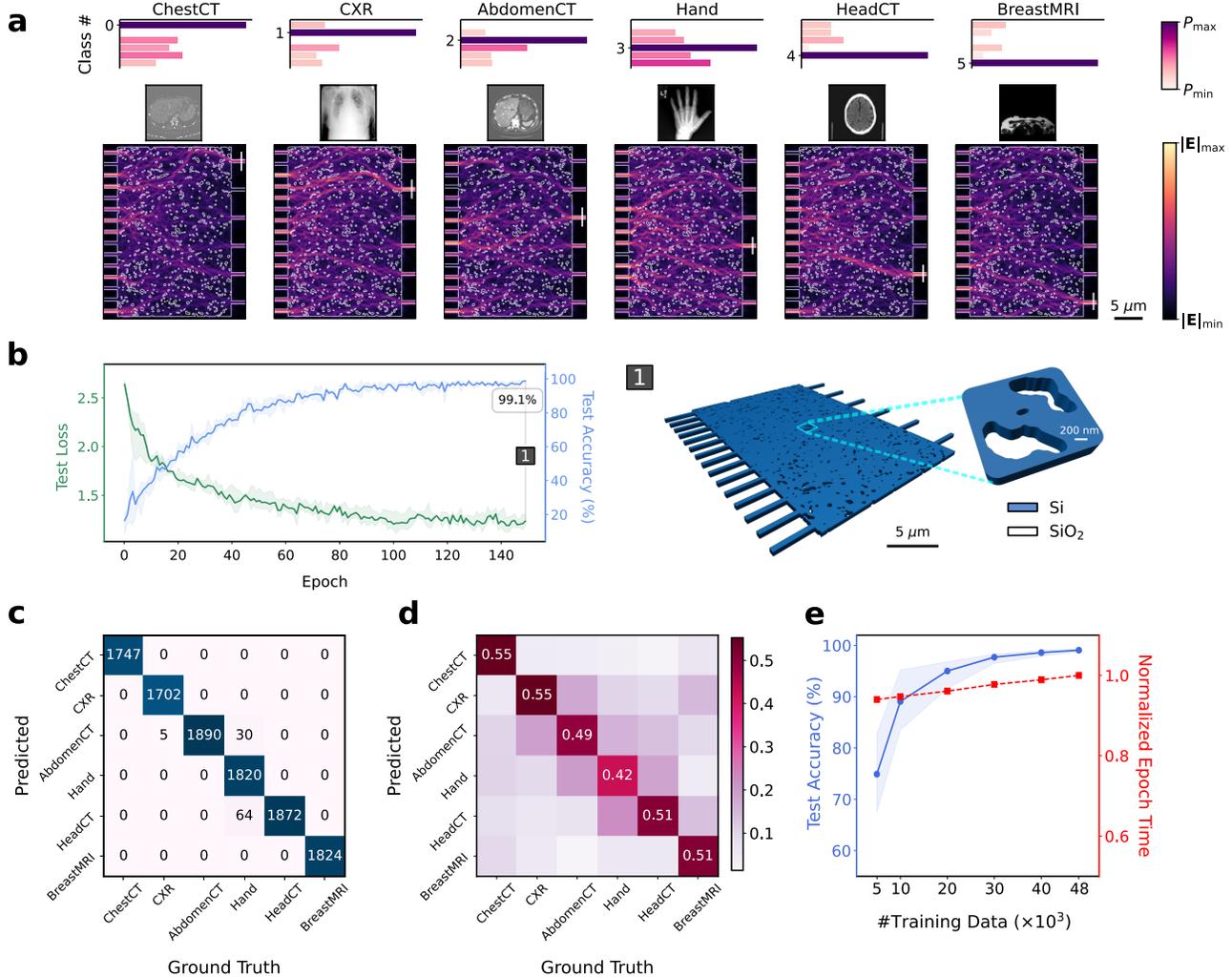

**Fig. 3 MedNIST 6-class classification result. a** Normalized power distribution $P$ with respect to a maximum value $P_{max}$ and minimum value $P_{min}$ across 6 random test class samples as measured at output ports, and corresponding E-field magnitude distributions at center-line with maximum and minimum values denoted as $|\mathbf{E}|_{max}$ and $|\mathbf{E}|_{min}$ respectively at 150th epoch. **b** Test loss and accuracy curves as a function of epochs with range overlayed across three random data permutations. Green: Test CE loss. Blue: Test classification accuracy. Inset 1: 3D-rendered schematic with B-spline interpolated topology at final epoch. **c** Confusion matrix for 6-class classification. **d** Energy density matrix at final epoch. **e** Blue: Mean test accuracy as a function of training data, with range overlaid across three random data permutations. Red: Max-normalized time taken for single training epoch as a function of training data.

**Hardware performance and inverse-design computational scalability.** Figure 4 shows the benchmark results for a single simulation job including key FDTD simulation parameters: mesh time, FDTD time and total wall time for MNIST and MedNIST datasets. The inverse-design process is conducted across three GPU platforms: Nvidia RTX 5090, RTX 4090, and V100. It is observed that the FDTD time for V100 is approximately 64 seconds to complete a single simulation for MNIST, compared to 48 seconds for RTX 4090 and 30 seconds for RTX 5090. The wall-clock time for a single simulation also accounts for factors such as mesh time - typically CPU bound. Moreover, given the size of the MedNIST chip is 1.5 times larger than MNIST, this increased footprint is reflected in the wall-clock time which is 1.46 times longer than MNIST.

The cumulative wall-clock time for the inverse-designed PNN is defined as the product of the single wall-clock time, the fixed number ($N + C$) 3D-FDTD simulations per epoch, and the total number of epochs. The total simulation time accounts for both the cumulative wall-clock time and total time required each epoch to derive spatial fields across the entire dataset. Our approach enables the linear separability of each simulation, allowing individual simulations to be independently executed on a clustered computing network[40]. This capability supports scheduling via makespan minimization, a strategy that optimally balances computational loads across available resources[41], significantly reducing overall compute time through parallelization. The inset of Fig. 4 illustrates the total simulation time under three distributed computing schemes. On a single GPU Linux node equipped with an Intel Core i7-processor, 128 GB of RAM, and a Nvidia RTX 5090 (32 GB) graphics card, the inverse-design optimization for MNIST PNN runs for approximately 29.7 hours. Using the same hardware configuration, the MedNIST PNN optimization requires approximately 56.3 hours. Combining RTX 4090 and RTX 5090 nodes reduces these times to 19.7 hours (MNIST) and 37.9 hours (MedNIST). Incorporating an additional V100 node further reduces the total simulation time to 17.1 hours for MNIST and 33.3 hours for MedNIST. These benchmarking results clearly demonstrate the effectiveness of our approach in leveraging parallel computing.

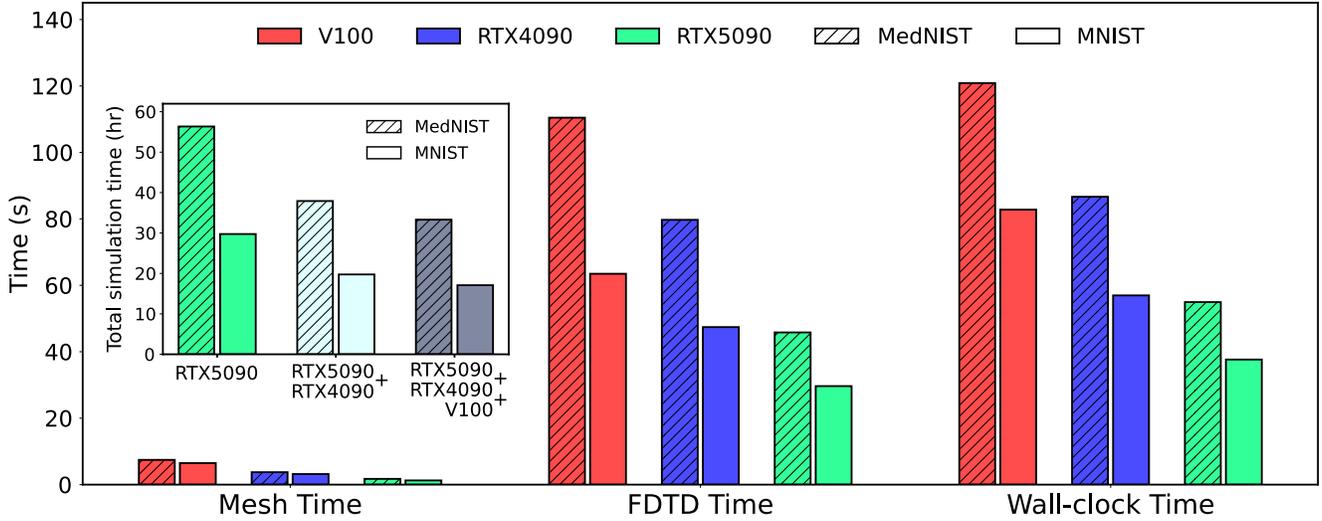

**Fig. 4 Benchmark timing results for a single simulation job and total simulation time.** Results are with respect to three computing nodes: Nvidia V100 with Xeon E5-2686 v4 CPU, RTX 4090 with Intel Core i9-10980XE CPU and RTX 5090 GPU with Intel Core i7-14700KF CPU. The bar chart compares single simulation mesh time, 3D-FDTD time, and wall-clock time for MNIST and MedNIST PNN designs. Inset compares the total simulation time under 3 distributed computing conditions: Single RTX 5090 GPU node, RTX 5090 + RTX 4090 GPU nodes, RTX 5090 + RTX 4090 + V100, demonstrating the simulation scalability via parallel processing.

**Fabrication and measurement.** To experimentally demonstrate the feasibility of our designs, we fabricate PNN accelerators for MNIST and MedNIST datasets on the SOI platform (see **Methods**). Figure 5a and 5b show the scanning electron microscope (SEM) images of the fabricated $20 \times 20\ \mu m^2$ MNIST and $30 \times 20\ \mu m^2$ MedNIST PNN accelerators, respectively, highlighting their compact footprints. Both input and output waveguides are designed to support the fundamental transverse-electric (TE) mode at 1550 nm, with a width of 500 nm and height of 220 nm. The minimal feature size across the device is 80 nm.

Optical signals from a CW laser source at 1550 nm are vertically coupled into the chip via vertical grating couplers (VGCs), distributed through a power splitter, and encoded with amplitude and phase information for input data preparation, and then processed via the inverse-designed PNN. To illustrate full device integration, Fig. 5c shows the MedNIST PNN chip wire-bonded to a printed circuit board (PCB) (See **Methods**) as the electrical interface through which optical amplitude and phase is adjusted for input data preparation.

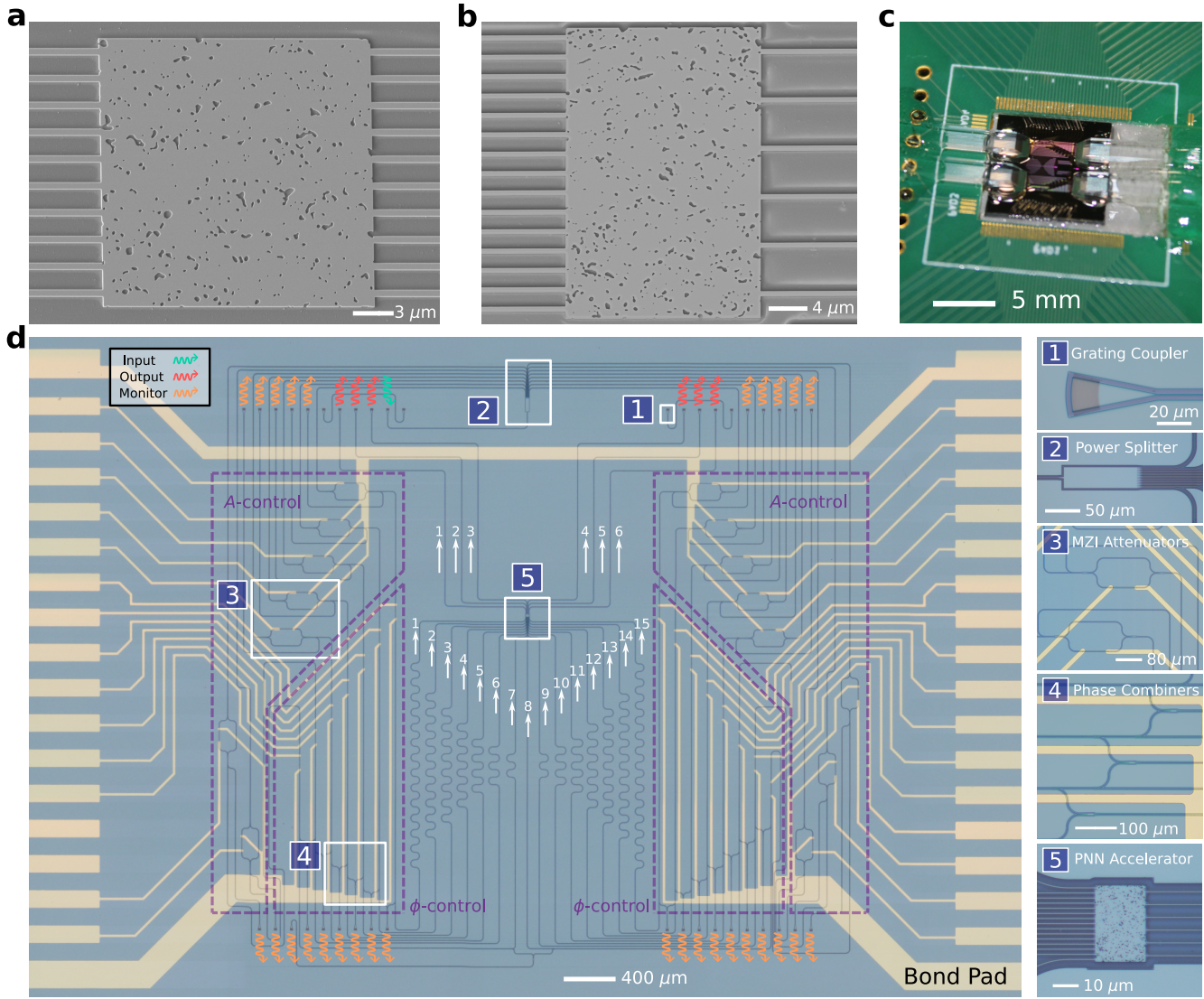

**Fig. 5 PNN accelerators fabricated on the SOI platform.** SEM images of **a** $20 \times 20$ μm² MNIST PNN with 10 input and 10 output waveguides and **b** $30 \times 20$ μm² MedNIST PNN with 15 input and 6 output waveguides. **c** Packaged SOI chip wire-bonded to PCB. **d** Microscope image of MedNIST PNN chip illustrating amplitude ($A$) and phase ($\phi$) control components for input data preparation, with optical input, output and monitoring ports. Gold traces and bond pads facilitate the interface between the photonic chip and MCU. Insets: Microscope images of (1) VGC, (2) 1×15 optical power splitter, (3) MZI-attenuator with optical monitoring ports, (4) optical phase combiners and (5) MedNIST PNN accelerator.

A detailed microscope image of the MedNIST chip is shown in Fig. 5d, clearly illustrating the VGC (Inset 1) for light coupling in and out of the chip, the multimode-interferometer (MMI) power splitter (Inset 2), Mach-Zehnder interferometers (MZIs) (Inset 3) and phase-combiners (Inset 4) for input data preparation, along with the compact PNN accelerator region (Inset 5) and monitoring ports. These monitoring ports enable direct measurement of optical power at critical locations within the device, facilitating real-time calibration. Finally, the optical power distribution from six output ports of the PNN accelerator are simultaneously measured using a multiport optical power meter to obtain classification results. The MNIST PNN chip follows a similar layout and packaging.

For experimental data preparation, we measure analogue optical input power to the PNN accelerators across 100 MNIST and 60 MedNIST test samples, where 10 samples are selected from each class. To verify the accuracy of these inputs, we measure a global mean average error (MAE) of 0.0277 for MNIST and 0.0310 for MedNIST with respect to the simulated inputs. These low MAE values confirm a strong correlation with the simulated amplitude-encoded features, validating the accuracy of our optical data preparation process.

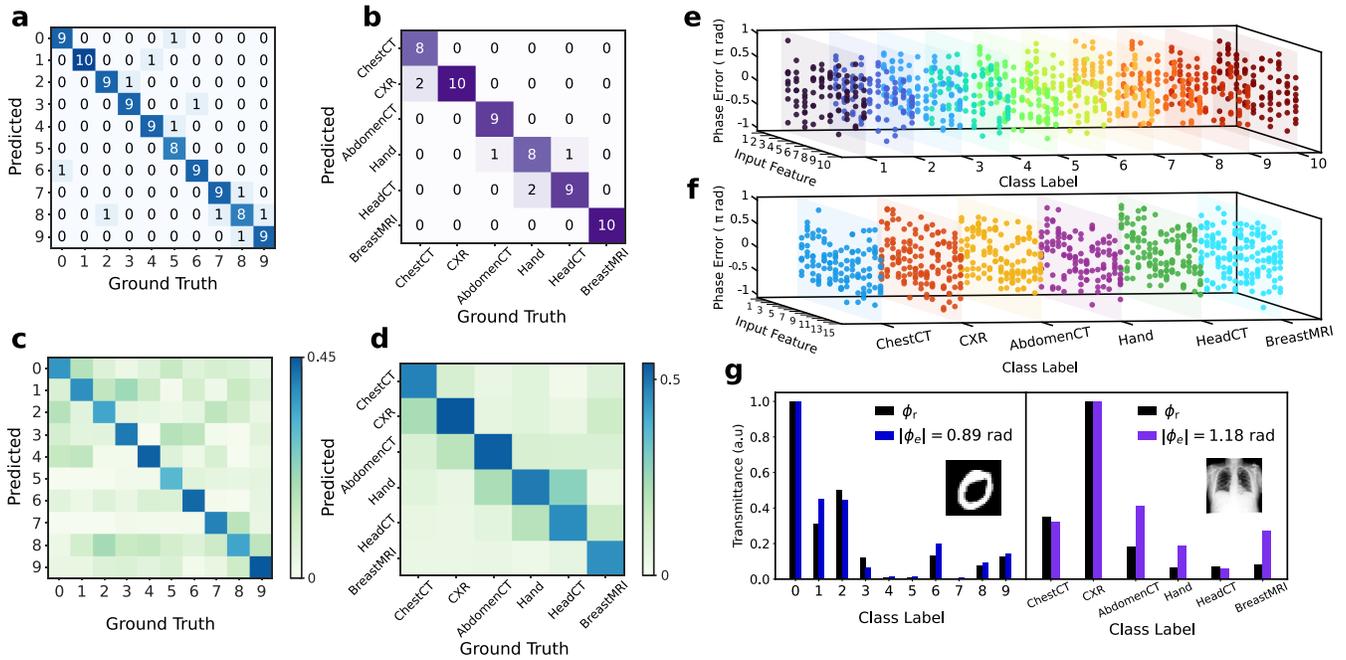

**Fig. 6 Experimental classification performance of inverse-designed PNN accelerator.** Confusion matrix for **a** (MNIST) and **b** (MedNIST) datasets. Energy density matrix for **c** (MNIST) and **d** (MedNIST) datasets. 3D plots of correctly classified data samples under a randomly induced phase variation between $[-\pi, \pi]$ radians for **e** 10 inputs MNIST and **f** 15 inputs MedNIST. **g** Comparison of transmission under mean absolute phase-error $|\phi_e|$ conditions of 0.89 radians (MNIST) and 1.18 radians (MedNIST) with respect to the reference phase $\phi_r$ for a random test-sample image from MNIST and MedNIST datasets respectively. $|\phi_e|$ is defined as the mean absolute deviation of the input phases from their ideal reference values across all channels.

For MNIST, an overall classification accuracy of 89% is achieved, with Class '1' exhibits the highest classification accuracy of 100%. For the MedNIST dataset, we achieve an overall classification accuracy of 90%, with the 'BreastMRI' and 'CXR' classes reaching the highest classification accuracy of 100%. Moreover, the MedNIST energy density matrix reveals an average of 50.5% of total output power concentrated at the correct output port. Among the correctly classified output ports, our results indicate only a 16.1% variation in maximum output power, demonstrating consistent classification performance across a higher-dimensional sequence of data. The classification score can be further improved by minimizing the power nonuniformities among the VGCs through enhanced fiber-to-chip alignment tolerance[42].

**Robustness verification.** To evaluate the robustness of the PNN devices experimentally, we explore the effect of input data phase deviations on classification performance (see **Methods**). Figures 6e and 6f show scatterplots of the maximum phase variation while still maintaining a correct classification for MNIST and MedNIST datasets respectively. Despite large input phase swings, the classification accuracy remains robust and consistent across both datasets. This robustness is primarily attributed to the amplitude-dominated encoding scheme, where classification decisions are governed predominantly by the strongest optical intensity features, reducing sensitivity to phase fluctuations. Figure 6g shows representative examples from the MNIST and MedNIST datasets under different mean absolute phase deviation conditions of 0.89 and 1.18 radians respectively, clearly illustrating the dominant energy distribution at the correct output ports despite these phase deviations.

**Discussion**

In this paper, we demonstrate ultra-compact inverse-designed PNN accelerators for on-chip image classification, based on a scalable, 3D-FDTD inverse-design methodology. By harnessing the inherent linearity of Maxwell's equations, our approach achieves computationally efficient and highly parallelizable optimization, enabling effective handling of increased dataset sizes. Experimental results on two SOI-chips reveal classification accuracies of 89% (MNIST) and 90% (MedNIST) in small footprints of $20 \times 20$ $\mu m^2$ and $30 \times 20$ $\mu m^2$ respectively. These results highlight the promise of inverse-designed PNN in emerging analog optical computing. Specifically, such designs, assisted by surrogate pre-processing, or in place of a digital fully connected output layer before classification, help alleviate the computational burden on electronic hardware. Looking ahead, our proposed framework readily scales to accommodate increased numbers of input channels, benefiting from its intrinsic parallelizability and computational efficiency. This scalability opens opportunities for processing higher dimensional and larger-scale datasets. Moreover, information encoded on multiple frequencies can be co-designed on the FDTD platform to enable multi-modal processing on a single device. Finally, our approach can be integrated into existing reconfigurable photonic architectures, offering a unified framework that bridges the gap between in-situ training capabilities and passive mathematical operations in the photonic domain.

## Methods

**Fabrication informed inverse-design parametrization.** At each inverse-design optimization step, we consider the parametrization of a latent feature space into a fabrication feasible, binarized material index distribution. These latent parameters vary continuously between 0 and 1, which correlates to $SiO_2$ ($\epsilon_{min}$) and Si ($\epsilon_{max}$) respectively for SOI waveguide. The feature space is randomly sampled from a uniform distribution then low pass filtered using a Gaussian blur kernel[43]. To embed fabrication constraints into the design, we first employ a low-pass filter via a conic kernel[34]. This filtered latent space is projected onto a binary distribution using the modified hyperbolic tan function[29] and is fully-binarized to enforce a real material permittivity matrix. The material is biased more toward Si ($\eta < 0.5$), where $\eta$ is the material biasing parameter. A B-spline approximation is utilized to further enforce minimum feature size and radial constraints in the graphic data system (GDS) design file, while facilitating a spatially adaptive resolution and anisotropic topology - subject to considered placement of control points[17]. A discrete contour which defines the boundary between material and void regions is first constructed using the Suzuki-Abe algorithm[44], followed by small feature removal using the *skimage-morphology* package in Python. The B-spline curve itself is composed of a linear combination of basis functions in 2D, expressed using the Cox-de Boor recursion formula[32, 33]. By tailoring both the degree of the curve $k$ and the number of knots, the density of approximated curves is effectively relaxed, thereby informing minimum feature and curvature constraints. In our case, a cubic ($k = 3$) B-spline approximation maps the discrete material voxels to a topology with a minimum feature size of 80 nm. Open-source Python library *gdspy* is used to translate these control points to create a GDS file at each design epoch.

**Numerical simulation setup.** To simulate the optical field distribution, a full-wave 3D-FDTD method (ANSYS Lumerical) is used. A uniform mesh is chosen with step size less than $\lambda/30$ to be equivalent in dimension to the parameter space, where $\lambda$ is the operating wavelength, and a non-uniform mesh is used outside the design region. A lightweight, semi-supervised autoencoder is used to learn a low-dimensional latent representation of the input data, which is then normalized and encoded onto the amplitude of light at 1550 nm. Open-source python framework *cupy* is used to compute and store the 128-bit complex-fields each of size $[n_x, n_y, n_z]$) in GPU's temporary memory. For both designs, a mini-batch size of 250 training samples is chosen to smoothen the gradient distribution and mitigate oscillatory behavior in optimization updates. The latent weights are randomly initialized and Adam[45] is used as the stochastic gradient optimizer, with a learning rate of 0.005 and exponential decay rates of 0.667, and 0.9 for the first and second-moment estimates respectively.

**Device fabrication and prototyping.** Both the MNIST and MedNIST PNN accelerators are fabricated via electron beam lithography with a foundry compatible minimum feature size (80 nm)[46] on a standard SOI wafer, which has a 220 nm thick top silicon layer sitting on top of a 2 μm thick buried oxide layer, above a 725 μm thick silicon substrate. VGCs[36,47] are fabricated with an etch depth of 70 nm. Plasma enhanced chemical vapor deposition process is used to deposit a 1 μm layer of $SiO_2$ onto the fabricated devices to act as an insulation layer between the electrodes and the device, which is sufficiently thick to minimize excess absorption due to the metal electrodes[48]. To facilitate thermo-optic tuning, thin-film titanium microheaters and gold traces are fabricated atop the $SiO_2$ cladding layer via a maskless lithography system (Heidelberg MLA100). For packaging and electrical interfacing, the devices are mounted onto a custom-designed PCB, which utilizes electroless nickel electroless palladium immersion gold contacts to provide the wire bonding material interface and electrical connectivity to a central MCU. The photonic chip is affixed to a thermoelectric cooler to ensure thermal stabilization. Temperature regulation is achieved via a thermistor embedded near the chip and connected through plated vias to the PCB.

**Data preparation, calibration and phase robustness testing.** Input light at 1550 nm from a tunable laser (Keysight N7778C) is fiber-coupled into the chip. A set of 10 and 15 MZIs are used for MNIST and MedNIST PNN devices respectively to facilitate amplitude control as data preparation. Each MZI is composed of a pair of 50:50 MMI splitters, integrated tapers, and S-bends with 30 µm arc radius. Downstream of MZIs, asymmetric 10:90 MMI are employed to tap a small portion of optical power for monitoring the attenuation in each channel. To monitor optical phase, adjacent channels are similarly tapped using 10:90 MMIs, and the extracted signals are recombined through 50:50 MMI combiners. The phase difference between channels is inferred from the combined optical power, where maximized combined optical power corresponds to phase equalization. For both MNIST and MedNIST chips, 10 representative samples from each class in the simulated test set are selected. The corresponding input channels are attenuated using MZI based amplitude control, with feedback from the optical power meter, while simultaneously monitoring and equalizing the relative phase between channels. For both amplitude and phase calibration, the voltage generated by a digitial-to-analog converter is swept between 1.1 V and 5 V with increments of 3 mV. The optimal voltage is identified by minimizing the error between the ground-truth and experimental optical transmission at 1550 nm. For phase robustness testing, the phase monitoring ports are used to identify the optimal phase corresponding to the highest classification result

– this phase is then served as a reference. The applied phase variation relative to this reference is inferred from the electrical power delivered to the thermo-optic phase shifters, based on their known tuning characteristics[49].

**Data availability.** The data that supports the findings of this study are available from the corresponding authors on reasonable request.

**Acknowledgements.** We thank S. Desai for discussions and assistance on device simulation, A. See for assistance with wire-bonding, L. Farar for assistance with device characterization and M. Kwok for assistance with device testing. The authors acknowledge the facilities as well as the scientific and technical assistance of the Research and Prototype Foundry Core Research Facility at the University of Sydney, a part of the Australian National Fabrication Facility. We also acknowledge the Sydney Informatics Hub for providing cloud computing resources, and the Royal Society of New South Wales for the Bicentennial Postgraduate Scholarship. J.S., G.L. and D. M. acknowledge support from Research Training Program Scholarships from the University of Sydney. This work was supported by the Sydney Research Accelerator Fellowship.

**Author contributions.** X. Y., L.L, S.S and J.S conceived the experiment. L. L. and S.S. fabricated the devices. J. S. and D. M. performed device modelling and simulations. G.L. and J.S. performed HPC cluster acceleration for device simulation. J.S., S.S and L.L carried out the device measurement and characterization. J.S., S.S., and X. Y. wrote the manuscript with contributions from all authors. All coauthors contributed to discussions of the protocol and results. X. Y. supervised the project.

**Correspondence** and requests for materials should be addressed to X. Y.